\documentstyle[prl,aps]{revtex}
\begin{document}

\twocolumn[{
\draft

\widetext
\title{Tunneling Spectroscopy of Quantum Charge Fluctuations in the Coulomb
Blockade}
\author{K. A. Matveev,$^{(1)}$ L. I. Glazman,$^{(2)}$ and H. U.
  Baranger$^{(3)}$}
\address{$^{(1)}$Massachusetts Institute of Technology, 12-105,
Cambridge MA 02139\\
$^{(2)}$Theoretical Physics Institute, University of
Minnesota, Minneapolis MN 55455\\
$^{(3)}$AT\&T Bell Laboratories,
600 Mountain Ave. 1D-230, Murray Hill NJ 07974}
\date{Submitted to Phys. Rev. Lett., April 25, 1995}

\maketitle
\vspace{-0.2in}

\mediumtext
\begin{abstract}
We present a theory of Coulomb blockade oscillations in tunneling through a
pair of quantum dots connected by a tunable tunneling junction.  The
positions and amplitudes of peaks in the linear conductance are directly
related, respectively, to the ground state energy and to the dynamics of
charge fluctuations. We study analytically both strong and weak interdot
tunneling. As the tunneling decreases, the period of the peaks
doubles, as observed experimentally. In the strong tunneling limit, we predict
a striking power law temperature dependence of the peak amplitudes.
\end{abstract}
%\pacs{PACS numbers: 73.20.Dx, 73.40.Gk}
\bigskip

}]

\narrowtext

The charge of an isolated conductor is quantized in units of the elementary
charge $e$. Surprisingly, even if the conductor is connected to a particle
reservoir by a tunnel junction, its charge can still be almost quantized at
low temperatures, a phenomenon known as the Coulomb
blockade \cite{GrabDev,Kastner}.
The simplest system which shows a Coulomb blockade consists of
a small metallic grain separated from a bulk lead by a thin
dielectric layer. An electron tunneling through the layer inevitably charges
the grain, thus increasing its energy by $E_C=e^2/2C$, where $C$ is the
capacitance of the grain. At temperatures $T\ll E_C$ a negligible
fraction of the electrons in the lead have an energy of order $E_C$, and
one might expect that no tunneling into the grain is possible. More
careful consideration shows, however, that even at $T=0$ the electrons can
tunnel to the virtual states in the grain, thus lowering the ground state
energy of the system\cite{Glazman}. Due to this virtual tunneling,
the average charge of the grain is no longer
quantized and acquires a correction proportional to the
conductance of the barrier. Charge quantization is completely destroyed
when the conductance of the barrier approaches $e^2/h$
\onlinecite{Flensberg,Matveev,Schoen}. Unfortunately a direct measurement of
the equilibrium properties, such as the average grain charge, comprises a
challenging, though not impossible\cite{two}, experiment.

Several recent experiments \cite{Waugh,Molenkamp,Delft,Stuttgart,Munich}
have probed some equilibrium properties of a Coulomb blockade
system by measuring the tunneling conductance through a
pair of coupled quantum dots. Focusing on the experiment of
Waugh, et al. \cite{Waugh}, we develop in this paper a quantitative
theory of the linear conductance in such a system; in particular, we
predict the gate voltage and temperature dependence of the
conductance.

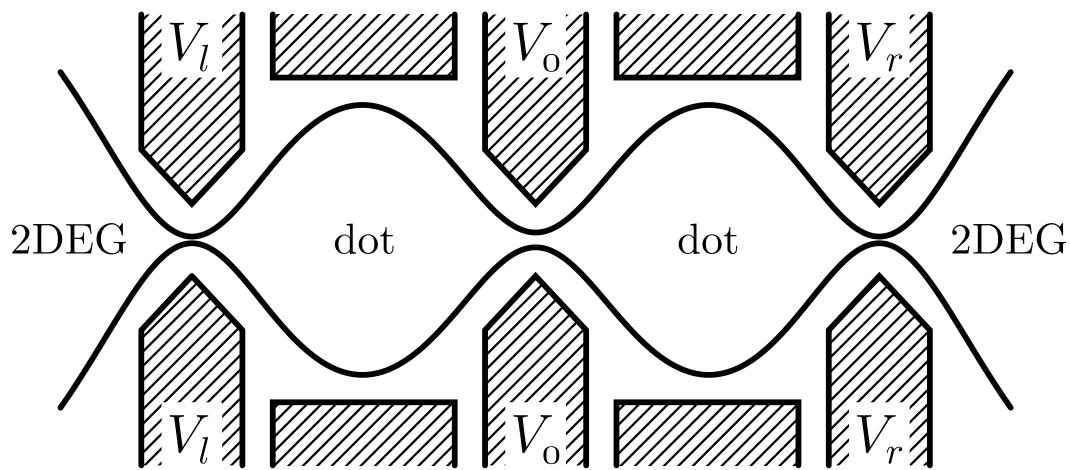
\begin{figure}
\caption{Schematic view of the double quantum dot system. The dots are
formed by applying negative voltage to the gates (shaded); the solid line
shows the boundary of the 2D electron gas (2DEG).
$V_l$ and $V_r$ create tunnel barriers between the dots and the leads
while $V_0$ controls the transmission coefficient through the constriction
connecting the dots.}
\label{fig:1}
\end{figure}

We will study the properties of a two-dot system which is
schematically shown in Fig.~\ref{fig:1}. The electrostatic energy of this
system is a quadratic form of three variables: the charges of each dot,
$eN_1$ and $eN_2$, and the gate voltage $V_g$. In the most
general case, this energy can be written in the following form:
\begin{eqnarray}
U(N_1,N_2) &=& E_C ( N_1 + N_2 - 2X)^2 \nonumber \\
&&+ \tilde{E}_C [ N_1 -N_2 +\lambda (N_1 + N_2) - \alpha X]^2 .
\label{U}
\end{eqnarray}
Here $X$ is a dimensionless parameter proportional to $V_g$, and the constants
$E_C$, $\tilde{E}_C$, $\lambda$, and $\alpha$ are determined by the geometry
of the system \cite{constants}.
In Eq.~(\ref{U}) we expressed the energy in terms of the
total number of particles in the two dots $N_1 + N_2$ and the
relative number $N_1 - N_2$. These variables are
convenient because the former is constant in the absence of
tunneling into the leads, and the latter describes charge fluctuations
between the dots. In this paper we will concentrate on the case of
symmetric geometry of the system, corresponding to $\lambda=\alpha=0$,
which is apparently the case in the experiment\cite{Waugh}.

We first discuss the location of the peaks in the conductance $G$ of the
double dot system.  In the limit of very small inter-dot conductance,
$G_0\ll e^2/h$, the peaks in $G$ occur when the electrostatic energy is
degenerate; that
is, when $U(n+1,n)$ equals either $U(n,n)$ or $U(n+1,n+1)$ where $n$ is an
arbitrary integer. As a result we find peaks at the following sequence of
gate voltages:
\begin{equation}
  \label{positions}
  X^* =
n+\frac{1}{2}\pm\frac{1}{4}\left(1-\frac{\tilde E_C}{E_C}\right).
\end{equation}
Weak electrostatic coupling between the dots corresponds to $E_C-\tilde
E_C\ll E_C$. In this case the two peaks with the same $n$ in sequence
(\ref{positions}) merge. This limit is observed in the
experiment\cite{Waugh}.

If all the junctions shown in Fig.~\ref{fig:1} have small conductances,
then the charges of both grains are well defined. In order to study the
effect of quantum fluctuations on the ground state energy of the system,
it is enough to increase only the inter-dot conductance $G_0$, keeping all
other conductances small. Under these conditions, the sum ${\cal N} \equiv
N_1 + N_2$ is constant and can still be treated as a $c$-number, but
$N_1 - N_2$ starts to fluctuate. These fluctuations change the
ground state energy, denoted $E_{\cal N}$, from the electrostatic estimate.
The peaks in $G$ now occur at gate
voltages $X$ where $E_{\cal N}(X)=E_{{\cal N}+1}(X)$. In order to find
$E_{\cal N}$, one should consider the quantum mechanical problem with the
Hamiltonian
\begin{equation}
H=H_0+H_T+E_C({\cal N}-2X)^2+
4{\tilde E}_C\!\left(\hat N_1 -\frac{{\cal N}}{2}\right)^2\!.
\label{H}
\end{equation}
Here the terms $H_0$ and $H_T$ describe, respectively, free electrons in
the dots and tunneling between the dots; we have replaced $N_2$ by
${\cal N} - N_1$, and $N_1$ should be treated as a quantum operator $\hat N_1$.
Typically the size of the
dots exceeds the effective Bohr radius ($\sim 100$\AA\ for
GaAs), and therefore the level spacing for electrons in the dots
is much smaller than the charging energy. We will neglect the level spacing
and assume a {\it continuous} spectrum in $H_0$, in contrast to
Refs.~\onlinecite{Sarma,Klimeck}. In the continuous model, the non-interacting
part of the Hamiltonian, $H_0+H_T$, does not depend on the total number of
particles ${\cal N}$.

In the Hamiltonian (\ref{H}) the parameter ${\cal N}$ is an integer.
However, formally we can consider Eq. (\ref{H}) at any ${\cal N}$,
enabling us to relate the ground state energy $E_{\cal N}$ to the average
value ${\bar N}_1({\cal N})$ of the first dot's charge:
\begin{equation}
\frac{\partial E_{{\cal N}}}{\partial{\cal N}}=2E_C({\cal N}-2X)-
4\tilde E_C \left[ {\bar N}_1({\cal N}) -\frac{{\cal N}}{2} \right] .
\label{partial}
\end{equation}
The condition for the peak position, $E_{{\cal N}+1}- E_{\cal N}=0$, can now
be obtained by integration of Eq.~(\ref{partial}). The result is
\begin{equation}
X^* = \frac{{\cal N}}{2} + \frac14
  -\frac{\tilde E_C}{E_C}\int_{\cal N}^{{\cal N}+1}
  \left[{\bar N}_1({\cal N}') -\frac{{\cal N}'}{2}\right] d{\cal N}'
\label{condition}
\end{equation}
where now ${\cal N}$ is an integer again.
In the limit $G_0\to 0$, the average ${\bar N}_1({\cal N'})$ is the
integer closest to ${\cal N'}/2$, and taking ${\cal N}= 2n$ and
${\cal N}=2n+1$ in Eq.~(\ref{condition}) reproduces Eq.~(\ref{positions}).
The advantage of Eq.~(\ref{condition}) is that it
is valid at any $G_0$. However, to make use of it, one has
to evaluate ${\bar N}_1({\cal N})$ at any ${\cal N}$, which is
a challenging quantum mechanical problem because of the Coulomb
interaction in Eq. (\ref{H}).
Fortunately, in the limit of a continuous spectrum, the Hamiltonian
(\ref{H}) coincides with
the one for a single dot connected by a tunnel junction to a massive lead.
The latter problem has been extensively studied in the limits of weak and
strong tunneling into the
dot\cite{Glazman,Flensberg,Matveev,Schoen,ZaikinSchoellerGrabert}.

For weak tunneling, $G_0\ll e^2/\hbar$, the deviation of $\bar N_1({\cal N})$
from an integer is small and can be found\cite{Glazman} from
second order perturbation theory in $H_T$. For ${\cal N}$ in the interval
$(2n-1, 2n+1)$, we get
\begin{equation}
  \label{perturbation}
  \bar N_1({\cal N})=n+\frac{\hbar G_0}{2\pi e^2}
                     \ln\frac{{\cal N}-2n+1}{2n+1-{\cal N}}.
\end{equation}
Substituting Eq.~(\ref{perturbation}) into Eq.~(\ref{condition}) and using
the periodicity of $\bar N_1({\cal N})$, we find peak
positions shifted by tunneling:
\begin{equation}
  \label{shifted}
  X^* \simeq n+\frac{1}{2}\pm\frac{1}{4}
    \left[1-\frac{\tilde E_C}{E_C}
            \left(1-\frac{4\ln2}{\pi}\frac{\hbar G_0}{e^2}\right)\right].
\end{equation}
The splitting of the two peaks with the same $n$ grows linearly with $G_0$.
The splitting here results from quantum charge fluctuations between the
dots, not changes in geometric capacitances as inferred in
Refs.~\onlinecite{Delft,Stuttgart}.

Clearly, the charge fluctuations grow with $G_0$. In the limit of strong
tunneling the discreteness of charge $N_1$ is no longer
important\cite{Flensberg,Matveev,Schoen}, and
$\bar N_1({\cal N})\to \frac{1}{2}{\cal N}$ [see Eq. (\ref{H})].
As a result the peaks are equidistant, $X^*=(2{\cal N} +1)/4$,
which is expected because in this limit the two
dots form a single conductor. The doubling of the period of the peaks from
this regime to that of Eq.~(\ref{positions}) is one of the main
observations of the experiment of Waugh, et al.\cite{Waugh}.

To find the peak positions as the system approaches the strong-tunneling
limit, one has to specify a model of the junction
between the dots. For electrostatically controlled dots in semiconductor
heterostructures, the junction is a microconstriction with smooth
boundaries\cite{Lesovik}. The ground state energy of such a system near
the strong-tunneling limit, when the reflection coefficient for the single
transverse mode propagating through the constriction is small ${\cal R} = 1 -
\pi\hbar G_0/e^2\ll1$, was found in Ref.~\onlinecite{Matveev} [Eq.~(48)].
Using that result, we get
\begin{equation}
  X^* \simeq \frac{2{\cal N} +1}{4} +
    (-1)^{\cal N} \frac{4e^{\bf C}}{\pi^3}\frac{\tilde E_C}{E_C}
    {\cal R}\ln\frac{1}{{\cal R}},
  \label{strong}
\end{equation}
where ${\bf C}\approx 0.5772$ is Euler's constant.

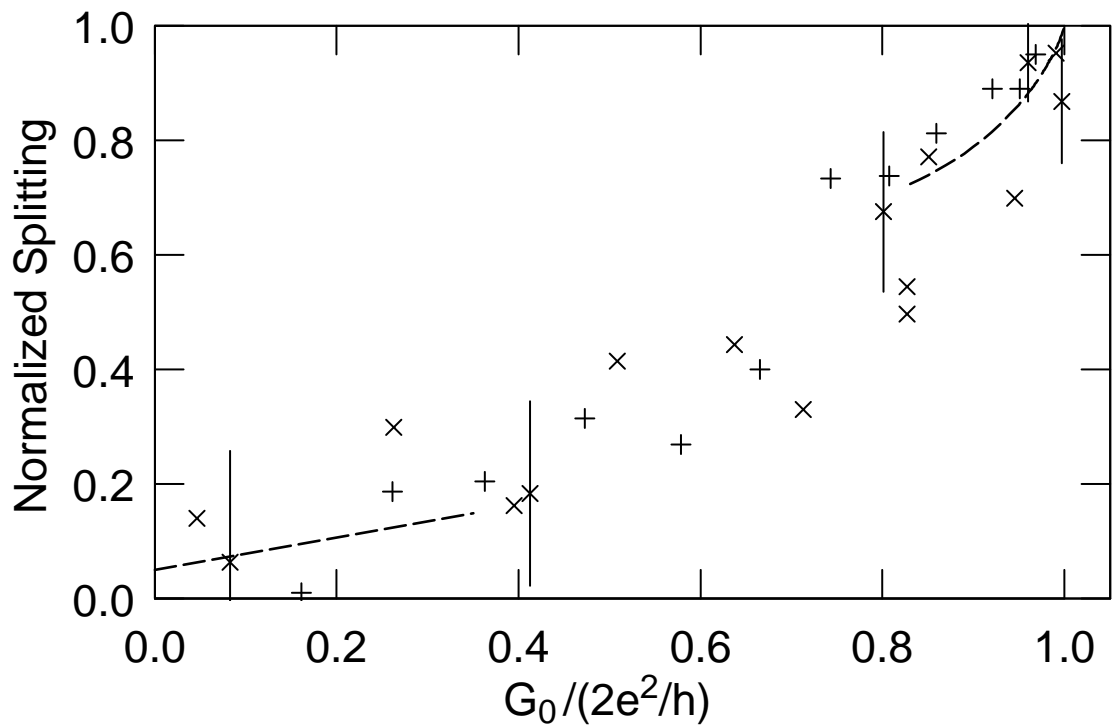
\begin{figure}
\caption{The normalized splitting of the Coulomb blockade peaks as a
function of the inter-dot conductance.  Our theoretical results
(dashed lines) are in good agreement with the experiment of Ref.
\protect\cite{Waugh} [$\times$ for data of Fig. 3(a) and $+$ for Fig.
3(b)].  The splitting is normalized by the period of the peaks in the
strong tunneling limit. A small asymmetry, $\alpha = 0.05$, has been
included based on the experimental parameters.}
\label{fig:2}
\end{figure}

Fig. \ref{fig:2} compares our results (\ref{shifted}) and
(\ref{strong}) to the observations of Waugh, et al.\cite{Waugh}.
Because the different gates are not independent, the relation between
the gate voltage $V_0$ at which the inter-dot conductance is measured
and the $V_0$ at which the splitting is measured is not known.  We
take this relation to be a rigid shift \cite{Waugh} and have used the
data of Ref. \cite{Waugh} with a shift of $-0.005\; V$. The agreement
between theory and experiment is very good indeed.

As we have seen, the {\it positions} of peaks in the linear conductance carry
information only about the ground-state energy of the two-dot system. To study
the excitations, one can analyze the {\it heights and shapes} of the peaks. If
the inter-dot tunneling is weak, $G_0\ll e^2/\hbar$, the excitation spectrum
consists of two independent quasiparticle spectra of the two dots. This
enables us to apply the standard master-equation technique\cite{GrabDev}
and find
\begin{equation}
G=\frac{G_lG_r}{2(G_l+G_r)}\frac{4E_C(X-X^*)/T}{\sinh[4E_C(X-X^*)/T]}.
\label{trivial}
\end{equation}
Here $X^*$ is the position of the center of a peak given by one of the
values in the sequence (\ref{shifted}).  In deriving Eq.~(\ref{trivial}) we
assumed
that the tunneling into the leads is much weaker than between the
dots, $G_l\sim G_r\ll G_0$. Thus Eq.~(\ref{trivial}) reproduces the result
for a single dot\cite{Shekhter}. This simple formula is valid only at
sufficiently low temperatures, when the width of a peak $\delta X\sim
T/E_C$ is much smaller than the spacing between the peaks. For small
capacitive coupling between the dots ($E_C \approx \tilde E_C$), this
yields $T\ll (\hbar G_0/e^2) E_C$.

In contrast to weak tunneling, at $G_0\sim e^2/\hbar$ the excitation spectra
of the two dots are not independent.
In this regime an electron tunneling into the left dot shakes up
the quantum state of the whole two-dot system, leading to a
suppression of the conductance at low temperature.
To illustrate this phenomenon, we calculate the temperature dependence of
the peak heights in the case of perfect inter-dot transmission,
$G_0=e^2/\pi\hbar$.

As we have seen, the conductance peaks in the strong-tunneling limit are
equally spaced, as if the double dot system were a single dot.
In fact, this remains true even for asymmetric double dots
[$\lambda, \alpha \neq 0$ in Eq.~(\ref{U})] because when ${\cal R} \to 0$
the energy is simply given by the first term in Eq.~(\ref{U}).
However, the specific geometry of the system---
whether it is one or two dots, and the degree of asymmetry---will
show up in the peak heights.  Unlike in a single dot,
the single-mode constriction impedes charge propagation between the two dots,
thus producing effects similar to those for a single junction coupled to
an environment\cite{GrabDev}.
When an electron tunnels from the lead into the left dot,
the other electrons in both dots must redistribute in order to minimize
the electrostatic energy: a charge
of $(1+\lambda)/2$ electrons must pass through the constriction. As a
result, the overlap of the two ground states, before and after the
tunneling, vanishes, as in Anderson's orthogonality catastrophe.

At non-zero temperature, the tunneling density of states is
suppressed as $T^\gamma$, where the exponent is related\cite{Schotte} to the
scattering phase shifts $\delta_m$ in each one-dimensional channel $m$ by
$\gamma=\sum_m (\delta_m/\pi)^2$. According to the Friedel sum rule,
$\delta_m/\pi$ is the average charge transferred into each channel. A
single-mode constriction provides two channels for each dot (two spins),
yielding $4$ channels in total.
In our case, $\delta_m/\pi=\pm(1+\lambda)/4$, where
the plus (minus) sign is for the channels in the right (left) dot.
Thus, the rate of tunneling into the left dot is
suppressed by the factor $T^{(1+\lambda)^2/4}$. For the rate
between the right dot and lead, one should replace
$\lambda$ by $-\lambda$. Because the junctions are connected in series,
the smaller of the two rates determines the conductance,
\begin{equation}
G=G_b
\left(\frac{T}{\tilde E_C}\right)^{\gamma}
F_{\gamma}\left(\frac{4E_C(X-X^*)}{T}\right),
\quad
\gamma=\frac{(1+|\lambda|)^2}{4}.
\label{peaks}
\end{equation}
Here the coefficient $G_b$ is of order $G_l\sim G_r$, and the peaks are
centered at $X^*=(2{\cal N}+1)/4$.
For a symmetric system, $\lambda=0$, the peak conductance obeys
$G\propto T^{1/4}$.
The temperature dependence in Eq.~(\ref{peaks}) can
be obtained analytically\cite{Baranger} in the spirit of the bosonization
approach\cite{Schotte,Matveev}; this technique also yields for the peak shapes
\begin{equation}
F_{\gamma}(x)=\frac{1}{\cosh(x/2)}
\frac{\left|\Gamma\left(1+\frac{\gamma}{2}+\frac{ix}{2\pi}\right)\right|^2}
{\Gamma(2+\gamma)}.
\label{shape}
\end{equation}
At $\gamma=0$ the shape given by
Eq.~(\ref{shape}) is identical to that for weak tunneling Eq.~(\ref{trivial}):
in both cases there is no charge transfer between the dots during the act of
tunneling, and the spectra of the two dots are decoupled.

Comparing our results in the weak and strong tunneling limits [Eqs.
(\ref{trivial}) and (\ref{peaks})], we see that
when $G_0$ grows the conductance of the system {\it decreases\/} due to
the orthogonality catastrophe.
For inter-dot conductances between these limits, the system
must crossover from temperature independent peak
heights for weak tunneling to the power-law suppression of the conductance
for strong tunneling. The theory of this crossover will be reported
elsewhere\cite{Baranger}.

We have seen that at $G_0=e^2/\pi\hbar$ an asymmetry of
capacitances, $\lambda\neq0$, causes temperature dependent peak heights
(\ref{peaks}). In this limit, the peaks are
equidistant and have the same height. We will now show that for weak
tunneling even a small asymmetry dramatically affects the whole pattern
of peaks. Below we study the asymmetry related to a
small non-zero $\alpha$ in the electrostatic energy
(\ref{U}); the results are easily generalized to include
a small $\lambda$ by replacing $\alpha\to\alpha + 2\lambda$.
Motivated by experiment\cite{Waugh}, we will assume $E_C=\tilde E_C$.

First, let us determine the positions of the peaks in conductance. The
condition for a peak is the degeneracy of the energy (\ref{U}) with
respect to adding an electron to either of the dots; this yields the two
sequences of peaks
\begin{equation}
  \label{sequences}
  X_1^*=\frac{n+1/2}{1-\alpha/2},
  \quad
  X_2^*=\frac{n+1/2}{1+\alpha/2}.
\end{equation}
Since $\alpha\ll 1$, the two periods are very close, and
one observes beats with an approximate superperiod of $\alpha^{-1}$.

The asymmetry lifts the degeneracy of the two states with an extra
electron on either the left or right dot. The energy gap between these two
states is \mbox{$\Delta (X)= 4\tilde E_C|\alpha X-m|$}, where $m$ is the
integer nearest to $\alpha X$. If the temperature is much lower that
$\Delta$, tunneling between two real states is suppressed. Instead, an
extra electron in the left dot can escape into the right lead through a
virtual state in the right dot. Because we assume that the level spacing is
small compared to temperature, the dominant escape mechanism is
{\it inelastic\/} co-tunneling\cite{Nazarov}, i.e., an extra
electron-hole pair is created in the right dot. The total
rate\cite{Nazarov} of such processes is proportional to $(T/\Delta)^2$
and limits the conductance at sufficiently
low temperatures. The calculation of peaks in conductance\cite{Baranger}
leads to
\begin{equation}
G\propto \left[\frac{T}{\Delta(X)}\right]^2
F_2\left(\frac{4E_C(X-X^*)}{T}\right).
\label{straightforward}
\end{equation}
Here the peak positions $X^*$ are defined by Eq.~(\ref{sequences}). The
oscillations of the energy gap $\Delta(X)$ modulate the
peak heights with the period $\alpha^{-1}$ in gate voltage $X$.

In conclusion, the fluctuation of electron charge between two
quantum dots strongly affects tunneling through such a structure.
First, the conductance peaks are split because of the lowering of the
{\it ground state energy} by charge fluctuations. This splitting,
and eventual halving of the period of the conductance peaks as the interdot
conductance grows, is a dramatic feature of the experimental data \cite{Waugh}
which is fully confirmed in our theory.
Second, the temperature dependence of the peak height and shape is
directly related to the {\it dynamics} of the quantum charge fluctuations.
For a double-dot connected by a reflectionless constriction,
this produces a striking fractional power law temperature dependence.
Our theory is valid for a wide range of temperatures, limited only by the
charging energy from above, and by the discrete energy level spacing from
below.

The work at MIT was sponsored by Joint Services Electronics Program Contract
DAAH04-95-1-0038 and at Minnesota by NSF Grant DMR-9423244.

\begin{references}
\bibitem{constants}
To be specific, the relations between our constants and the experimental
ones of Ref. \protect\cite{Waugh} are
$X = V_a (C_{2a} + C_{3a})/2e$,
$E_c = e^2 /2(C_2 + C_3)$,
$\tilde E_c = e^2 (C_2^{-1} + C_3^{-1})/8$,
$\lambda = (C_{3a}-C_{2a})/(C_2 + C_3)$, and
$\alpha = 2(C_{2a}-C_{3a})[(C_{2a}+C_{3a})^{-1}- (C_2+C_3)^{-1}]$,
where $a=5$ or $6$ throughout.
\end{references}
\end{document}